\documentclass[%
 reprint,
 amsmath,amssymb,
 aps,
 pre,
floatfix
]{revtex4-2}

\usepackage{graphicx}

\usepackage{dcolumn}
\usepackage{bm}

\begin{document}

\title{Modeling Tissue Detachment and Rupture Using an Extended Vertex Model with T2-inverse Transitions}


\author{Shota Nishimoto}
\affiliation{%
 Graduate School of Life Sciences, Ritsumeikan University, 1-1-1 Noji-higashi, Kusatsu, Shiga 525-8577, Japan
}%


\author{Yuichi Togashi}
 \email{togashi@fc.ritsumei.ac.jp}
\affiliation{%
 Graduate School of Life Sciences, Ritsumeikan University, 1-1-1 Noji-higashi, Kusatsu, Shiga 525-8577, Japan
}%
\affiliation{
 RIKEN Center for Biosystems Dynamics Research, 2-2-3 Minatojima-minamimachi, Chuo, Kobe, Hyogo 650-0047, Japan
}%

\date{\today}

\begin{abstract}
The vertex model is widely used to describe the mechanics of epithelial tissues, but its conventional formulation assumes that all cells remain tightly packed and always share edges with their neighbors, making it difficult to represent local detachment or gap formation. Here, we propose a minimal extension of the vertex model that enables cell detachment by introducing a new topological transformation, T2-inverse, which acts as the inverse of the classical T2 transition. When the imbalance of forces acting on a vertex, quantified by a tension metric $T$, exceeds a threshold, the T2-inverse splits the vertex into multiple vertices and creates a closed polygon that is incorporated as a pseudo-cell. This operation allows the model to represent the emergence and propagation of local detachment events. Using this framework, we simulate the stretching of a cell sheet and show that force-induced local detachments can accumulate to produce macroscopic tissue rupture. These results demonstrate that the proposed model extends the capability of the vertex model to describe tissue-level breakdown processes, including detachment and tearing, and provides a foundation for studying a broader class of epithelial mechanical phenomena.
\end{abstract}


\maketitle

\section{Introduction}

Epithelial tissues exhibit collective mechanical behaviors that arise from local interactions among tightly packed cells.
To describe such systems, Honda first introduced a polygonal representation of epithelial sheets in which cell shapes and mechanical forces are encoded through the positions of vertices, establishing the conceptual foundation of what is now known as the vertex model \cite{Honda1978,Nagai2001}.
Building on this framework, Farhadifar \emph{et al.} formulated the modern energy-based vertex model, in which cell area elasticity, perimeter contractility, and interfacial tension are combined into a unified mechanical potential \cite{Farhadifar2007}.
This formulation has since become the standard basis for computational studies of epithelial mechanics, providing a minimal yet powerful description of tissue-level behavior with single-cell resolution \cite{Staple2010,Fletcher2014}.

Epithelial vertex models have been applied to a wide range of morphogenetic processes in which dynamic changes in cell connectivity play essential roles.
An important feature of vertex models is their ability to represent such changes in connectivity through discrete topological transitions.
T1 transitions, in which a shrinking cell--cell interface is replaced by a new interface between a previously non-adjacent pair of cells, correspond to neighbor exchange and are closely related to the polarized cell rearrangements.
For example, T1-mediated neighbor exchanges have been shown to drive convergent extension during \textit{Drosophila} germ-band elongation \cite{Zallen2004}.
The same mechanism underlies chirality-induced shear flows and collective migration driven by anisotropic junctional tension \cite{Sato2015}.
The model has also been used to explain epithelial packing geometries \cite{Farhadifar2007}, cell extrusion during tissue overcrowding \cite{Marinari2012}, and fluidization-driven wound healing \cite{Tetley2019}.
Beyond two-dimensional tissues, vertex-based approaches have been extended to three-dimensional morphogenesis, including epithelial folding and invagination \cite{Honda2004, Okuda2015}.
Through these applications, the vertex model has become a central tool for linking cellular mechanics to emergent biological function.

Although the vertex model has been successfully applied to a wide range of epithelial processes, it retains a fundamental restriction that the cells form a confluent tiling in which every cell remains in continuous contact with its neighbors, and hence intercellular gaps or free boundaries cannot spontaneously emerge \cite{Farhadifar2007, Fletcher2014}.
As a consequence, the standard formulation cannot represent local detachment, partial loss of adhesion, or gap formation events that are essential for understanding tissue rupture and mechanical failure.
This limitation is particularly relevant because epithelial experiments increasingly demonstrate that local loss of tissue integrity can involve nucleation of holes or gaps \cite{Sonam2023, Lv2024}.

A recent extension introduces the T4 (detach) transition, in which a cell--cell interface is removed once the normal stress on the junction exceeds a critical threshold, allowing a finite gap to open in an otherwise confluent tissue \cite{Chen2022,Gao2025}.
This framework enables fracture-like behavior and has successfully demonstrated strain-rate-dependent plastic deformation and ductile-to-brittle transitions under mechanical stretching.

Nevertheless, it remains challenging to represent the initiation and progressive propagation of a local separation.
By definition, T4 transition produces cell--cell detachment as a single, abrupt event in which an entire interface fails at once.
Because the rule is defined at the level of edge duplication, it remains inherently two-dimensional and does not capture the asymmetric, progressive nature of many biological separation processes.

To overcome these limitations, we introduce a minimal extension of the vertex model that enables cell--cell detachment through a new topological operation, the T2-inverse transition.
Defined as the inverse of the classical T2 transition, this operation is triggered by a tension metric that quantifies local force imbalance at a vertex.
Unlike T4, which removes an entire interface in a single step, the T2-inverse transition allows detachment to nucleate asymmetrically and propagate gradually from one side of a junction.
Because the operation is formulated through local vertex creation rather than duplication of an edge that represents an entire two-dimensional interface, it also provides a route towards natural extension to three-dimensional geometries.
Together, these features provide a physically grounded framework for modeling the initiation and progression of epithelial separation, including tearing and rupture, that remain inaccessible in the conventional vertex model.

\section{Method}

\subsection*{Conventional vertex model}

\begin{figure}[htb]
  \centering
  \includegraphics[width=\linewidth]{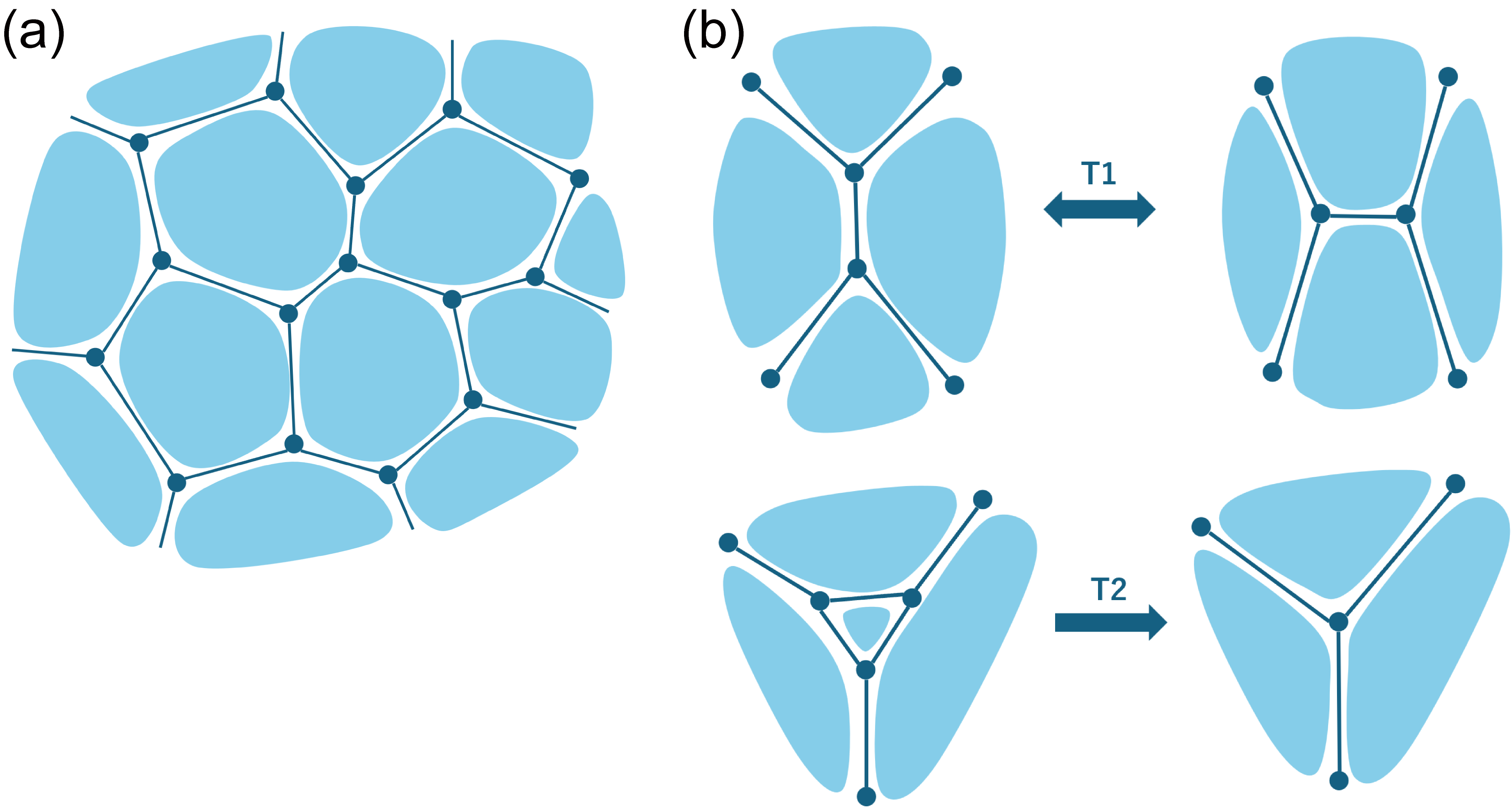}
  \caption{Overview of the conventional vertex model. (a) Concept of the vertex model. (b) T1 and T2 transitions.}
  \label{fig:vertex_conventional_overview}
\end{figure}

In the vertex model, each cell is represented as a polygon, and the entire tissue is described as a network of vertices connected by edges (Fig. \ref{fig:vertex_conventional_overview}(a)).
The mechanical state of the tissue is described by an energy functional consisting of area elasticity, perimeter elasticity, and interfacial tension terms. Vertex positions evolve according to overdamped dynamics driven by the gradient of the total energy,
\[
\eta \frac{d\mathbf{r}_v}{dt} = -\frac{\partial E}{\partial \mathbf{r}_v},
\]
with
\[
E = \sum_{cells} \frac{k_A}{2}(A - A_{0})^2 
  + \sum_{cells} \frac{k_P}{2}(P - P_{0})^2
  + \sum_{edges} k_{cc/ce} L.
\]
Here, \(A\) and \(P\) denote the area and perimeter of each polygonal cell,
and \(L\) represents the length of the cell--cell interface or cell--external-space interface. \(A_{0}\) and \(P_{0}\) are the preferred values of the area and perimeter \cite{Sato2015}.
In this study, $A_{0}$ is rescaled to $1$ without loss of generality.

In addition to the mechanical dynamics described above, the vertex model also incorporates topological rearrangements that modify the connectivity of the network (Fig. \ref{fig:vertex_conventional_overview}(b)).
Topological rearrangements are implemented through two standard operations. T1 transitions exchange neighboring cells when an edge becomes sufficiently short, and T2 transitions remove a cell whose area shrinks to zero.
These operations allow the conventional vertex model to capture neighbor exchanges in densely packed tissues but do not permit local detachment or gap formation.

\subsection*{T2-inverse: a new topological transformation}

\begin{figure}[htb]
  \centering
  \includegraphics[width=\linewidth]{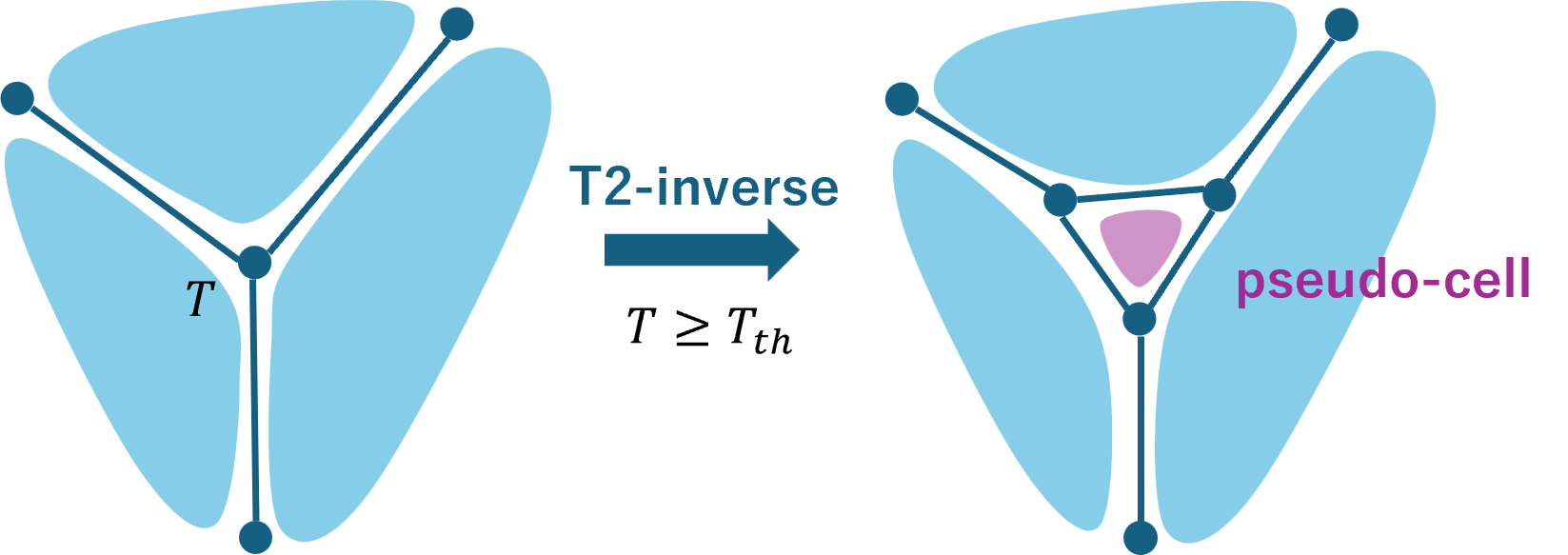}
  \caption{
    Conceptual illustration of the T2-inverse transition.
    A high-tension vertex undergoes a T2-inverse split,
    generating three vertices and a pseudo-cell that represents local gap formation.
  }
  \label{fig:T2_inverse_overview}
\end{figure}

To enable detachment, we introduce T2-inverse, a new topological operation that acts as the inverse of the classical T2 transition (Fig. \ref{fig:T2_inverse_overview}).
When triggered, T2-inverse splits a single vertex into three vertices and inserts new edges connecting them, thereby creating a closed polygonal region.
This region is incorporated into the system as a pseudo-cell, which is treated as an external space: it preserves the network connectivity but does not participate in the area or perimeter elasticity.
The operation opens a gap at the location of the split while preserving the polygonal representation of the surrounding cells.

\subsection*{Tension metric and detachment criterion}

\begin{figure}[htb]
  \centering
  \includegraphics[width=\linewidth]{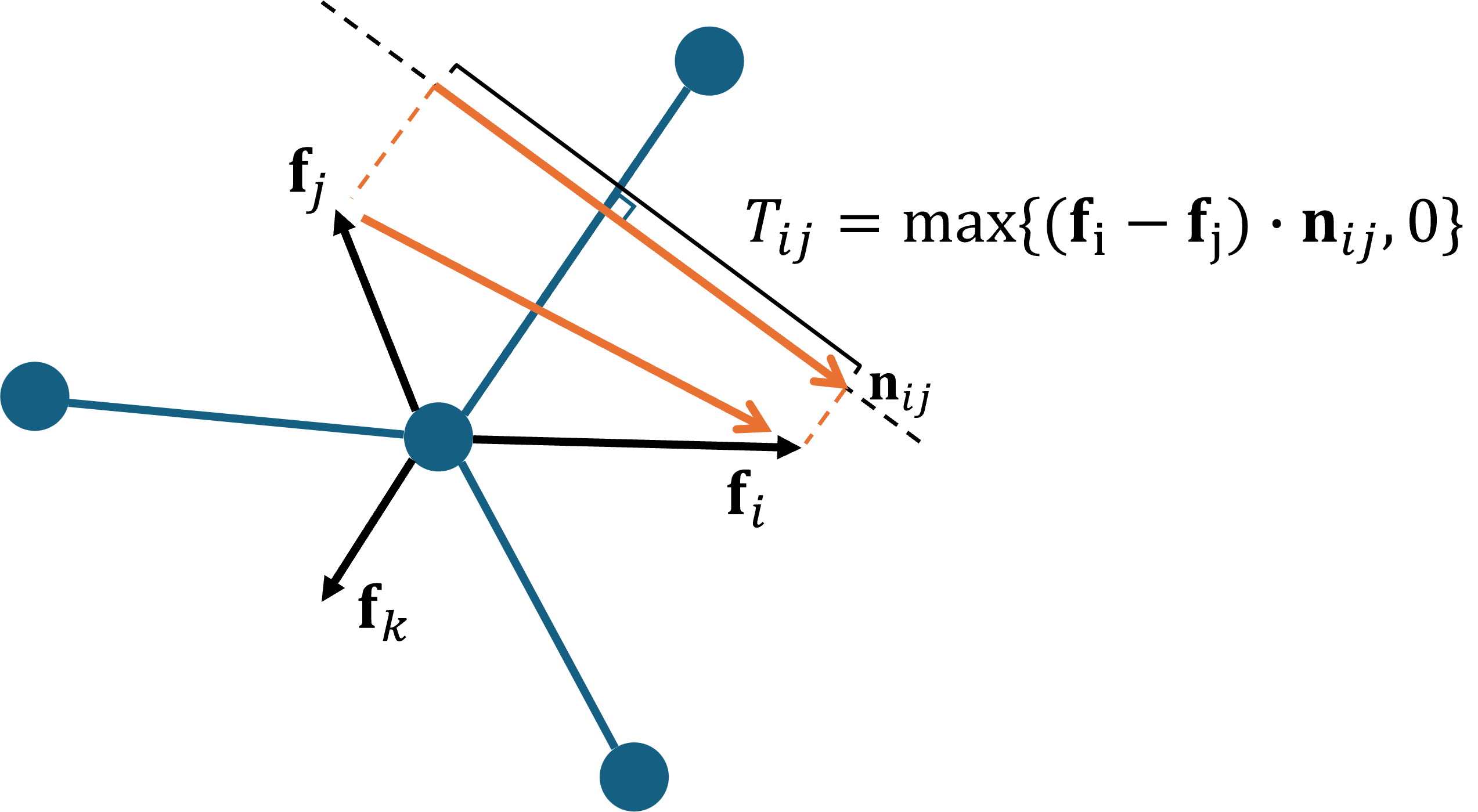}
  \caption{
    Geometric definition of the tension metric $T_{ij}$.
  }
  \label{fig:T_metric}
\end{figure}

Detachment is triggered based on the imbalance of forces acting on a vertex.
For a vertex shared by three cells $i$, $j$, and $k$, we define the force imbalance along the interface between cells $i$ and $j$ as
\[
T_{ij}=\max\lbrace(\mathbf{f_{\mathnormal{i}}}-\mathbf{f_{\mathnormal{j}}})\cdot \mathbf{n_{\mathnormal{ij}}}, 0\rbrace,
\]
where $\mathbf{f_{\mathnormal{i}}}$ and $\mathbf{f_{\mathnormal{j}}}$ are the forces exerted by the respective cells and $\mathbf{n_{\mathnormal{ij}}}$ is the unit normal vector of their shared edge (Fig. \ref{fig:T_metric}). The tension metric at the vertex is then defined as
\[
T=\max \{ T_{ij},T_{jk},T_{ki}\} 
.\]
When $T$ exceeds a threshold $T_{\mathrm{th}}$, the vertex is considered mechanically unstable, and a T2-inverse transition is initiated.

\subsection*{Detachment simulation}
To demonstrate the capabilities of the extended framework, we simulated tissue detachment under uniaxial stretching.
We initialized the system as a confluent sheet composed of regular hexagonal cells, each with unit area (Fig.~\ref{fig:speed_compare} (top)). Two types of initial configurations were considered: a uniform sheet and a sheet containing a small pre-existing crack.
First, the sheet was relaxed to the steady state (until the change in total energy fell below $10^{-6}$ per step), which corresponds to 0\% extension below (note that due to the line tension term $k_{cc/ce} L$, the area and perimeter decreased from the preferred values $A_{0}$ and $P_{0}$).
Then, uniaxial stretching was applied by moving the left and right boundaries outward at a constant speed. Two stretching speeds were tested ($0.1$ and $0.5$).
The top and bottom boundaries were left free, allowing transverse relaxation.

To evaluate the effect of the proposed T2-inverse operation, simulations were performed both with and without enabling T2-inverse.
When a gap formed, the region was treated as a pseudo-cell representing external space, preserving network connectivity without contributing to the mechanical energy.
All other simulation parameters were identical across conditions and are summarized in Table~\ref{tab:common_params}.

\begin{table*}
  \centering
  \caption{Common simulation parameters used across all conditions.}
  \begin{tabular}{llc}
    \hline
    Parameter & Description & Value \\
    \hline
    $A_0$ & Preferred cell area & $1$ \\
    $P_0$ & Preferred cell perimeter & $2\sqrt[4]{12}$ (perimeter of regular hexagon with area $1$) \\
    $k_A$ & Area elasticity coefficient & $20.0$ \\
    $k_P$ & Perimeter elasticity coefficient & $20.0$ \\
    $k_{cc}$ & Line tension for cell--cell interfaces & $1.0$ \\
    $k_{ce}$ & Line tension for cell--external space interfaces & $1.0$ \\
    $\eta$ & Damping coefficient (overdamped dynamics) & $1.0$ \\
    $\Delta t$ & Integration time step & $1.0 \times 10^{-4}$ \\
    $L_{th}$ & T1 threshold & $1.0 \times 10^{-3}$ \\
    $A_{th}$ & T2 threshold & $1.0 \times 10^{-5}$ \\
    $T_{th}$ & T2-inverse threshold & $10.0$ \\
    Boundary condition & Left/right moved, top/bottom free &  \\
    \hline
  \end{tabular}
  \label{tab:common_params}
\end{table*}

\section{Results}

\subsection*{Baseline behavior without T2-inverse}

\begin{figure*}
  \centering
  \includegraphics[width=0.75\linewidth]{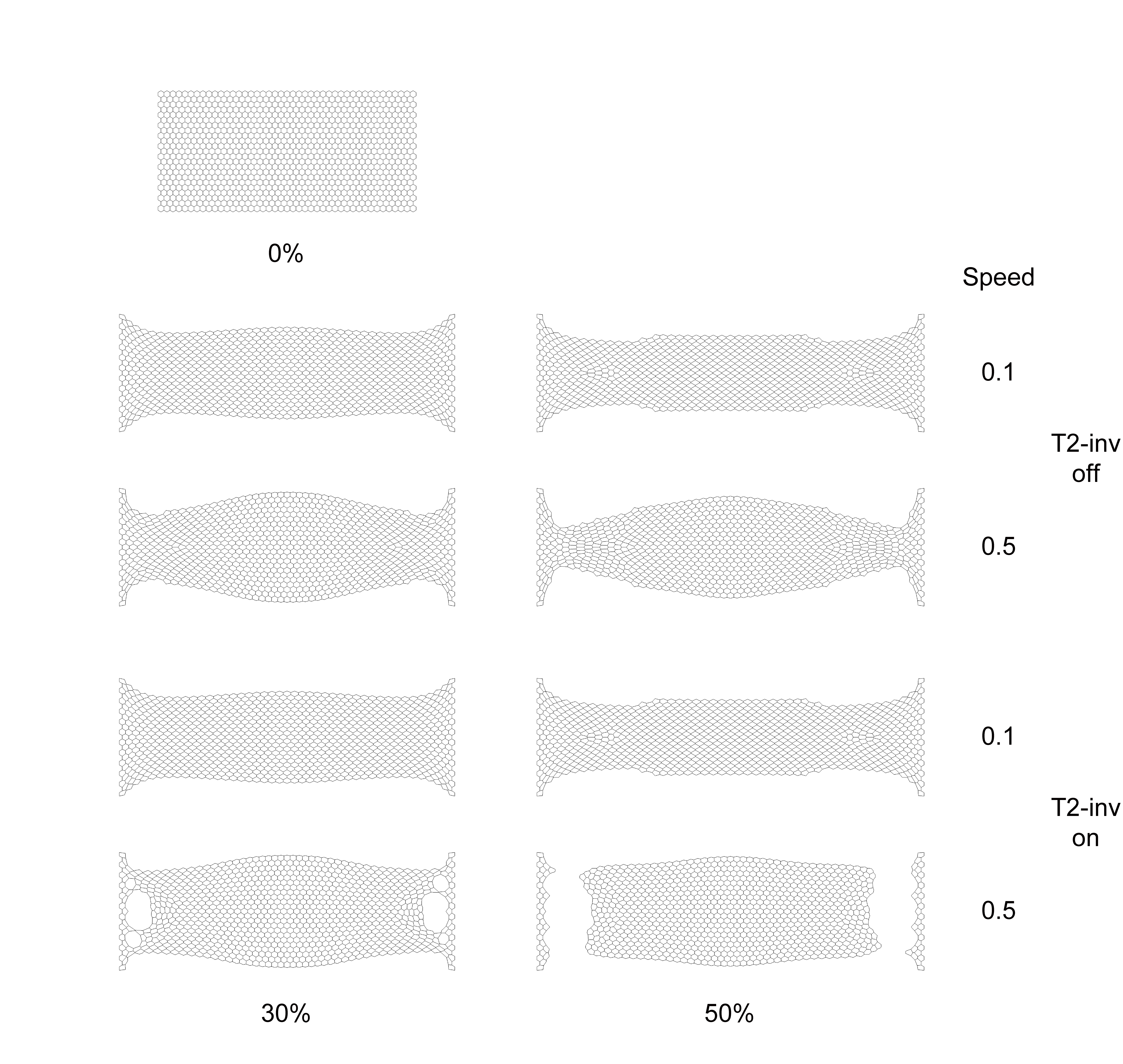}
  \caption{
    Snapshots of the sheet under uniaxial stretching, with and without T2-inverse transitions.
    Top: the initial configuration (0\% extension).
    Middle (T2-inv off): with T2-inverse disabled; at both stretching speeds (0.1 and 0.5),
    the sheet remains cohesive up to 50\% extension, and no gap formation is observed.
    Bottom (T2-inv on): with T2-inverse enabled; at the higher speed (0.5),
    T2-inverse events occur and gaps form, leading to detachment, whereas no gap is observed
    at the lower speed (0.1).
  }
  \label{fig:speed_compare}
\end{figure*}

\begin{figure}
  \centering
  \includegraphics[width=\linewidth]{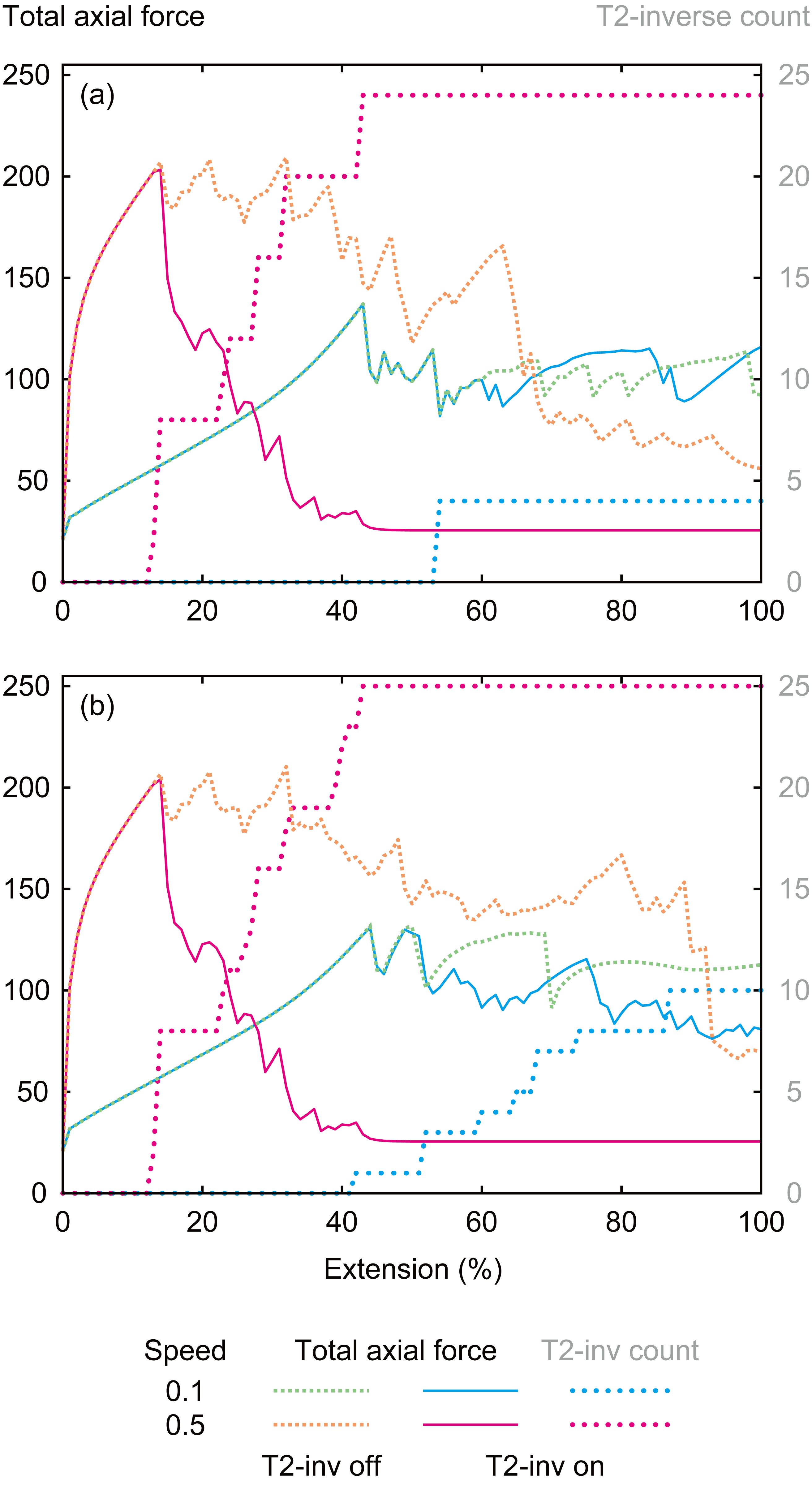}
  \caption{
    Force--extension curves for the systems (a) without and (b) with a pre-existing crack in the initial condition.
    The horizontal axis shows the percentage extension (100\% means twice the steady-state length),
    and the vertical axis shows the total axial force in the stretching direction (left axis).
    Dashed lines correspond to the cases without T2-inverse transitions (stretching speed 0.1 (green) and 0.5 (orange)),
    and solid lines correspond to those with T2-inverse transitions (stretching speed 0.1 (cyan) and 0.5 (magenta)).
    For the latter cases, the cumulative number of T2-inverse transitions is indicated by dotted lines in the same color (right axis).
    Note that, with T2-inverse at speed 0.5, most of the sheet detached from the ends at $\sim 43$\% extension, and the remaining force (constant $\sim 25$) after that point simply corresponds to the dragging friction without stretching.
  }
  \label{fig:force-extension}
\end{figure}

We first examined how the sheet responds to uniaxial stretching when T2-inverse is disabled.
Simulations were performed at two stretching speeds ($0.1$ and $0.5$), and snapshots at 0\%, 30\%, and 50\% extension are shown in Fig.~\ref{fig:speed_compare} (middle).
In both cases, cells elongated along the stretching direction while maintaining nearly constant area.
T1 transitions occurred as deformation progressed, but these neighbor exchanges did not lead to interface collapse or gap formation.
The sheet remained cohesive up to 50\% extension at both speeds.

These results show that stretching speed alone does not induce detachment when T2-inverse is disabled.
Although the tissue undergoes topological rearrangements through T1 transitions, these do not produce the type of topological change required for gap formation.
This baseline behavior provides a reference for evaluating how T2-inverse alters the mechanical response under identical loading conditions.

\subsection*{T2-inverse enables detachment under fast stretching}

We next examined the effect of enabling T2-inverse.
At the lower stretching speed ($0.1$), no T2-inverse events occurred.
Although T1 transitions were observed, the tension metric $T$ at each vertex remained below the activation threshold, and the sheet stayed cohesive throughout the deformation.

At the higher speed ($0.5$), T2-inverse events appeared during stretching.
Fig.~\ref{fig:speed_compare} (bottom) shows snapshots at 0\%, 30\%, and 50\% extension. Fast stretching generated localized increases in vertex tension, and once the metric $T$ exceeded the threshold, T2-inverse was triggered.
This event produced a gap, which subsequently expanded and led to detachment.

These results indicate that T2-inverse induces failure only under rapid deformation, where vertex tension becomes sufficiently large to activate the operation.

\subsection*{Effect of stretching speed on the spatial distribution of vertex tension}

\begin{figure*}
  \centering
  \includegraphics[width=\linewidth]{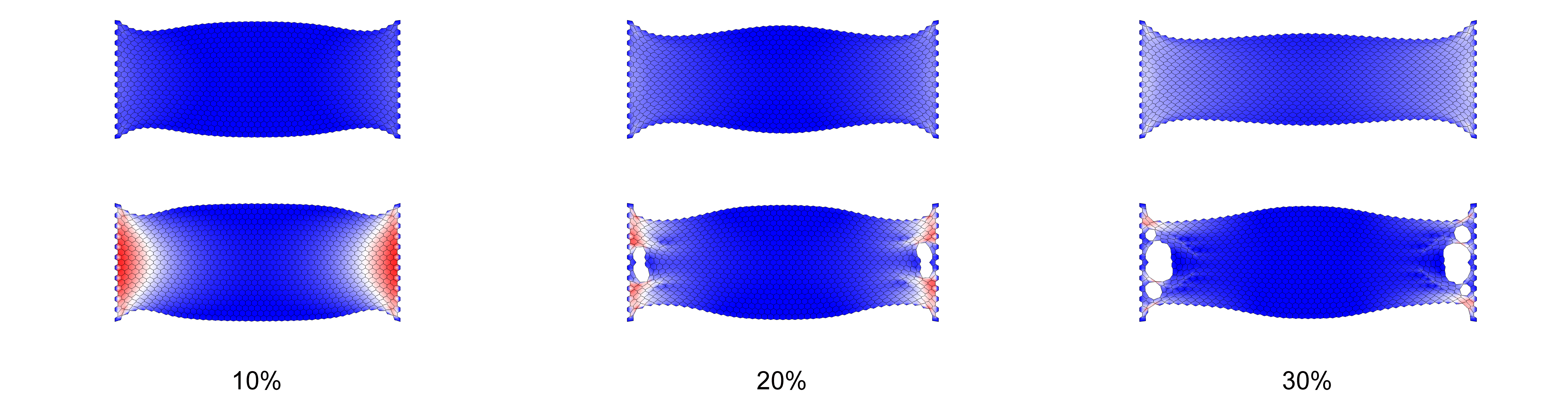}
  \caption{
    Spatial distribution of the tension metric $T$ during stretching. Top row: stretching speed $0.1$. Bottom row: stretching speed $0.5$. 
    At the higher speed, localized regions of elevated tension appear as deformation progresses, approaching the activation threshold for
    T2-inverse around 30\%. (Color scale: blue corresponds to $T = 0.0$ and red to $T = 10.0$.)
  }
  \label{fig:tension_speed_compare}
\end{figure*}

To examine how stretching speed influences the buildup of mechanical tension, we analyzed the spatial distribution of the tension metric $T$ during deformation. Fig.~\ref{fig:tension_speed_compare} shows $T$ at 10\%, 20\%, and 30\% extension for speeds $0.1$ and $0.5$.

At the lower speed ($0.1$), the tension remained relatively uniformly low across the sheet, and no localized peaks were observed.
In contrast, at the higher speed ($0.5$), regions of elevated tension emerged as deformation progressed.
These high-tension regions triggered the activation of T2-inverse.
The difference in behavior is also observed in the force--extension curves and the cumulative number of T2-inverse shown in Fig. \ref{fig:force-extension}.

These results indicate that stretching speed strongly affects the spatial pattern of vertex tension, and that rapid deformation generates localized tension peaks that can exceed the activation threshold for T2-inverse.

\subsection*{Effect of pre-existing cracks under slow stretching}

\begin{figure*}
  \centering
  \includegraphics[width=\linewidth]{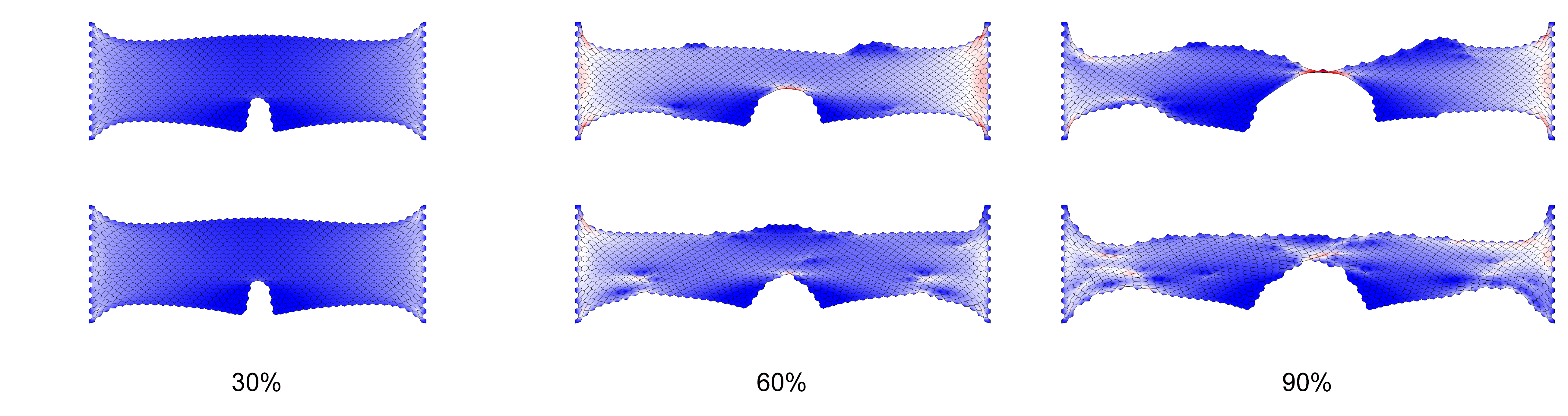}
  \caption{
    Tension distribution around a pre-existing crack under slow stretching (speed $0.1$).
    Top row: T2-inverse OFF. Bottom row: T2-inverse ON, where the crack-tip
    tension triggers a local split and the crack extends naturally. (Color scale: blue corresponds to $T = 0.0$ and red to $T = 10.0$.)
    }
  \label{fig:defect_compare}
\end{figure*}

To assess how pre-existing defects influence detachment, we introduced a small crack at the boundary of the sheet and examined the resulting tension distribution during slow stretching ($0.1$) (Fig. \ref{fig:defect_compare}).
When T2-inverse was enabled, tension accumulated at the crack tip and exceeded the activation threshold, triggering local separation from the crack tip and allowing the crack to extend along an irregular, tortuous path.

In contrast, when T2-inverse was disabled, the accumulated tension was relieved only through repeated T1 rearrangements.
As a result, failure occurred more rapidly but along an unnaturally straight fracture line that did not follow the crack tip.
This behavior was qualitatively different from the crack-driven propagation observed with T2-inverse.

These results indicate that the introduction of T2-inverse not only enables detachment to initiate from pre-existing cracks, but also produces more natural-looking fracture patterns, whereas models without T2-inverse tend to generate artificially straight failure boundaries driven by T1 transitions.

\section{Discussion and Conclusion}

We introduced T2-inverse, a new topological operation that enables local gap formation in vertex-based tissue models.
Unlike the conventional vertex model, which relies solely on T1 rearrangements and therefore cannot represent natural detachment, T2-inverse allows high-tension vertices to split and form a local gap when the tension metric exceeds a threshold.

Through controlled stretching experiments, we showed that detachment is strongly dependent on deformation speed.
In homogeneous sheets, slow stretching produced relatively uniformly low tension, whereas fast stretching generated localized tension peaks that triggered T2-inverse and led to detachment.
Pre-existing cracks further modified this behavior: under slow stretching, tension accumulated at the crack tip and activated T2-inverse, allowing the crack to extend.
Under fast stretching, however, failure occurred away from the crack before tension could concentrate at the defect.

Although similar strain-rate-dependent behavior was observed with T4 \cite{Gao2025}, the propagation of detachment is represented differently.
Recent experiments have revealed physical routes to tissue failure \cite{Casares2015, Bonfanti2022}, and also observed stress-relaxation and loading-rate-dependent behavior \cite{Khalilgharibi2019, Esfahani2021, Prakash2021, Duque2024, Arora2025}, which may be further compared with these simulations.
Another advantage over T4 is the extensibility to three-dimensional geometries, although a concrete formulation is still to be developed.

In conclusion, these results demonstrate that the T2-inverse transition provides a physically consistent mechanism for modeling detachment and crack propagation in vertex-based tissue simulations.
The operation captures both speed-dependent failure and defect-driven crack extension, enabling more realistic descriptions of tissue rupture.

\section{Acknowledgments}

We thank Masashi Fujii and Sonja Tarama for insightful comments.
This work was partially supported by JSPS KAKENHI Grant Number JP19H05424 (project \emph{Singularity Biology}).

\bibliographystyle{unsrt}
\bibliography{references}

\end{document}